\documentclass[aps,pra,reprint,superscriptaddress,amsmath,amssymb,floatfix]{revtex4-2}

\usepackage{graphicx} 
\usepackage{hyperref}
\usepackage{array}

\usepackage{soul, todonotes}
\usepackage{multirow}

\begin{document}

\title{Training neural networks with end-to-end optical backpropagation}

\author{James Spall}
\affiliation{Clarendon Laboratory, University of Oxford, Parks Road, Oxford,  OX1 3PU, UK}

\author{Xianxin Guo}
\email{Xianxin.Guo@physics.ox.ac.uk}
\affiliation{Clarendon Laboratory, University of Oxford, Parks Road, Oxford,  OX1 3PU, UK}
\affiliation{Lumai Ltd., Wood Centre for Innovation, Quarry Road, Headington, Oxford, OX3 8SB, UK}

\author{A. I. Lvovsky}
\email{Alex.Lvovsky@physics.ox.ac.uk}
\affiliation{Clarendon Laboratory, University of Oxford, Parks Road, Oxford,  OX1 3PU, UK}
\affiliation{Lumai Ltd., Wood Centre for Innovation, Quarry Road, Headington, Oxford, OX3 8SB, UK}

\date{\today}

\begin{abstract}
Optics is an exciting route for the next generation of computing hardware for machine learning, promising several orders of magnitude enhancement in both computational speed and energy efficiency. However, to reach the full capacity of an optical neural network it is necessary that the computing not only for the inference, but also for the training be implemented optically. The primary algorithm for training a neural network is backpropagation, in which the calculation is performed in the order opposite to the information flow for inference. While straightforward in a digital computer, optical implementation of backpropagation has so far remained elusive, particularly because of the conflicting requirements for the optical element that implements the nonlinear activation function. In this work, we address this challenge for the first time with a surprisingly simple and generic scheme. Saturable absorbers are employed for the role of the activation units, and the required properties are achieved  through a pump-probe process, in which the forward propagating signal acts as the pump and backward as the probe. Our approach is adaptable to various analog platforms, materials, and network structures, and it demonstrates the possibility of constructing neural networks entirely reliant on analog optical processes for both training and inference tasks.
\end{abstract}

\maketitle

\section*{Introduction}
Machine learning, one of the most revolutionary scientific breakthroughs in the past decades, has completely transformed the technology landscape, enabling innovative applications in fields ranging from natural language processing to drug discovery.
As the demand for increasingly sophisticated machine learning models continues to escalate, there is a pressing need for faster and more energy-efficient computing solutions. In this context, analog computing has emerged as a promising alternative to traditional digital electronics~\cite{roy2019towards,yao2020fully,xu202111,feldmann2021parallel,sludds2022delocalized,lin2018all,shen2017deep}. A particularly exciting platform for analog neural networks (NNs) is optics, in which the interference and diffraction of light during propagation implements the linear part of every computational layer~\cite{shastri2021photonics, Wu2022analog}.

Most of the current analog computing research and development is aimed at using the NN for inference~\cite{wetzstein2020inference, shastri2021photonics}. Training such NNs, on the other hand, is a challenge. This is because the backpropagation algorithm \cite{lecun2015deep}, the workhorse of training in digital NNs, requires the calculation to be performed in the order opposite to the information flow for inference, which is difficult to implement on an analog physical platform. Hence analog models are typically trained offline (\emph{in silico}), on a separate digital simulator, after which the parameters are transferred to the analog hardware. In addition to being slow and inefficient, this approach can lead to errors arising from imperfect simulation and systematic errors (`reality gap'). In optics, for example, these effects may result from dust, aberrations, spurious reflections and inaccurate calibration~\cite{spall2022hybrid}.

To enable learning in analog NNs, different approaches have been proposed and realized~\cite{Buckley2023}. Several groups explored various `hardware-in-the loop' schemes, in which, while the backpropagation was done \emph{in silico}, the signal acquired from the analog NN operating in the inference regime was incorporated into the calculation of the feedback for optimizing the NN parameters~\cite{wright2022deep,spall2022hybrid,zhou2021large,tam1990learning,mennel2020ultrafast,li2018efficient,wang2019situ,feldmann2019all}. This has partially reduced the training error, but has not addressed the low speed and inefficiency of \emph{in silico} training.

Recently, several optical neural networks (ONNs) were reported that were trained online (\emph{in situ}) using methods alternative to backpropagation. Bandyopadhyay \emph{et al.}~trained an ONN based on integrated photonic circuits using simultaneous perturbation stochastic approximation, i.e.~randomly perturbing all ONN parameters and using the observed change of the loss function to approximate its gradient~\cite{bandyopadhyay2022single}. Filipovich~\emph{et al.}~applied direct feedback alignment, wherein the error calculated at the output of the ONN is used to update the parameters of all layers~\cite{Filipovich2022}. However both these methods are inferior to backpropagation as they take much longer to converge, especially for sufficiently deep ONNs~\cite{Bartunov2018}.

An optical implementation of the backpropagation algorithm was proposed by Hughes \emph{et al.}~\cite{hughes2018training}, and recently demonstrated experimentally~\cite{pai2023experimentally}, showing that the training methods of current digital NNs can be applied to analog hardware. However, their scheme omitted a crucial step for optical implementation: backpropagation through nonlinear activation layers. Their method requires digital nonlinear activation and multiple opto-electronic inter-conversions inside the network, complicating the training process. Furthermore, the method applies only to a specific type of ONN that uses interferometer meshes for the linear layer, and does not generalise to other ONN architectures. Complete implementation of the backpropagation algorithm in optics, through all the linear and nonlinear layers, that can generalise to many ONN systems, remains a highly challenging goal.

In this work, we address this long-standing challenge and present the first complete optical implementation of the backpropagation algorithm in a two-layer ONN. The gradients of the loss function with respect to the NN parameters are calculated by light travelling through the system in the reverse direction.
The main difficulty of all-optical training lies in the requirement that the nonlinear optical element used for the activation function needs to exhibit different properties for the forward and backward propagating signals. Fortunately, as demonstrated in our earlier theoretical work~\cite{guo2021backpropagation} and explained below, there does exist a group of nonlinear phenomena,
which exhibits the required set of properties with sufficient precision.

We optically train our ONNs to perform classification tasks, and our results surpass those trained with a conventional \textit{in silico} method. Our optical training scheme can be further generalized to other platforms using different linear layers and analog activation functions, making it an ideal tool for exploring the vast potential of analog computing for training neural networks.

\begin{figure*}
    \includegraphics[width=0.9\textwidth]{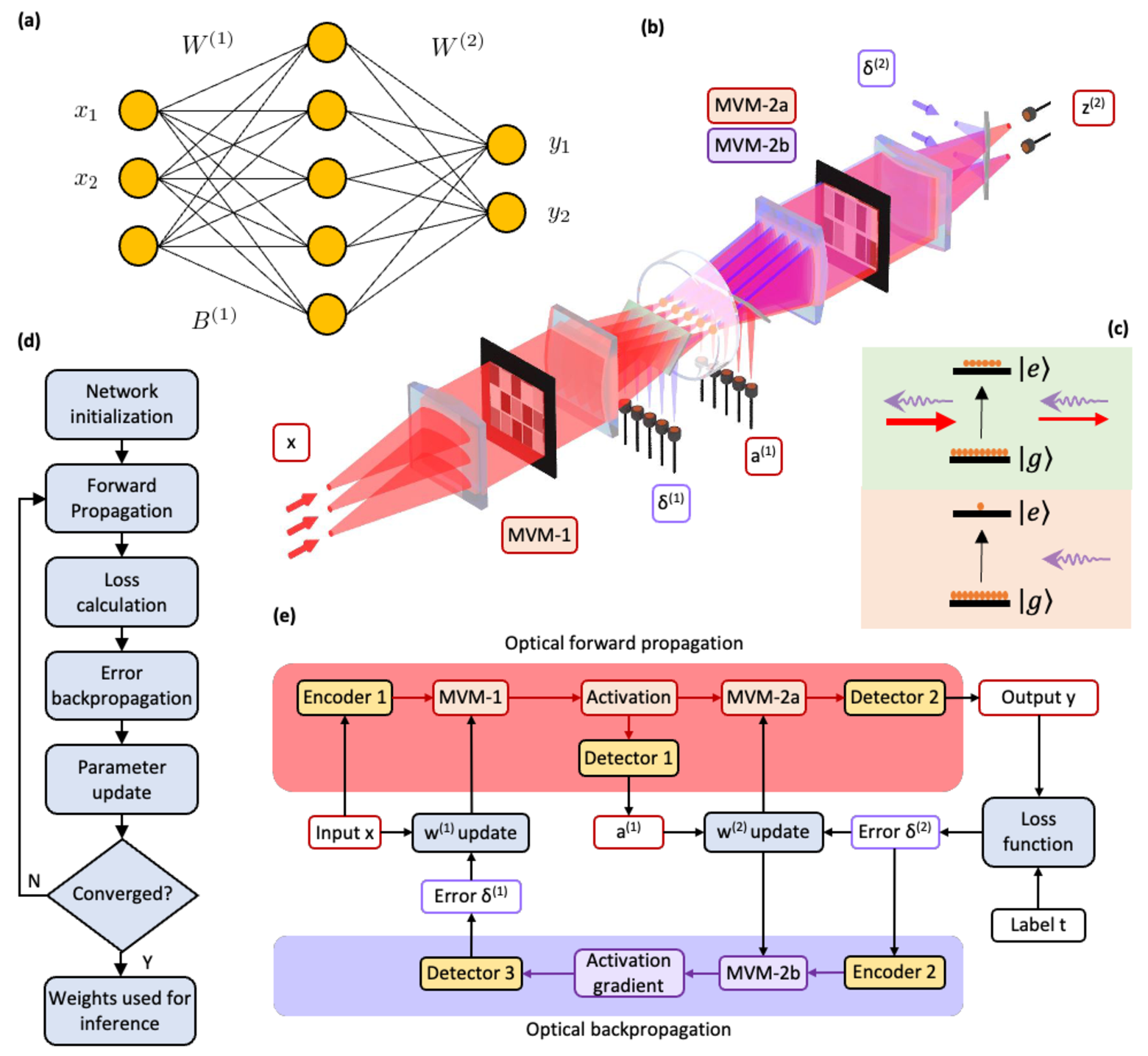}
    \caption{\textbf{Illustration of optical training.} \textbf{(a)} Network architecture of the ONN used in this work, which consists of two fully-connected linear layers and a hidden layer. \textbf{(b)} Simplified experimental schematic of the ONN. Each linear layer performs optical MVM with a cylindrical lens and a spatial light modulator (SLM) that encodes the weight matrix. Hidden layer activations are computed using SA in an atomic vapour cell. Light propagates in both directions during optical training. \textbf{(c)} Working principle of SA activation. The forward beam (pump) is shown by solid red arrows, backward (probe) by purple wavy arrows. The probe transmission depends on the strength of the pump and approximates the gradient of the SA function. For high forward intensity (top panel), a large portion of the atoms are excited to the upper level. Stimulated emission produced by these atoms largely compensates the absorption due to the atoms in the ground level. For weak pump (bottom panel), the excited level population is low and the absorption is significant. \textbf{(d)} Neural network training procedure. \textbf{(e)} Optical training procedure. Both signal and error propagation in the two directions are fully implemented optically. Loss function calculation and parameter update are left for electronics without interrupting the optical information flow.
    }
    \label{fig1}
\end{figure*}

\section*{Optical training algorithm}
We consider a multilayer perceptron --- a common type of NN which consists of multiple linear layers that establish weighted connections between neurons, interlaid by activation functions that enable the network to learn complex nonlinear functions. To train the NN, one presents it with a training set of labeled examples and iteratively adjusts the NN parameters (weights and biases) to find the correct mapping between the inputs and outputs.

The training steps are summarised in Fig.~\ref{fig1}(d). The weight matrices, denoted $W^{(i)}$ for the $i$-th layer, are first initialized with random values. Each iteration of training starts by entering the input examples from the training set as input vectors $x = a^{(0)}$ into the NN, and forward propagating through all of its layers. In every layer $i$, one performs a matrix-vector multiplication (MVM) of the weight matrix and the activation vector,
\begin{equation}
z^{(i)} = W^{(i)}\times a^{(i-1)},
\label{eq: mvm}
\end{equation}
followed by element-wise application of the activation function $g(\cdot)$ to the resulting vector:
\begin{equation}
a^{(i)} = g\left(z^{(i)}\right).
\label{eq: activation}
\end{equation}

The output $y=a^{(L)}$ of an $L$-layer NN allows one to compute the loss function $\mathcal{L}(y,t)$ that determines the difference between the network predictions $y$ and ground truth labels $t$ from the training set. The backpropagation algorithm helps calculating the gradient of this loss function with respect to all the parameters in the network, through what is essentially an application of the chain rule of calculus. The network parameters are then updated using these gradients and optimization algorithms such as stochastic gradient descent. The training process is repeated until convergence.

The gradients we require are given by~\cite{lecun2015deep} as
\begin{equation}\label{otimes}
    \frac{\partial \mathcal{L}}{\partial W^{(i)}} =  \delta^{(i)} \otimes a^{(i-1)},
\end{equation}
where $\delta^{(i)}$ is referred to as the ``error vector'' at the $i$th layer and $\otimes$ denotes the outer product. The error vector is  calculated as
\begin{align}
    \delta^{(i-1)} &= \left( {W^{(i)}}^T \times \delta^{(i)} \right)  g'\left( z^{(i-1)}\right),
\label{eq: backprop}
\end{align}
going through layers in reverse sequence. The expression for the error vector $\delta^{(L)}$ in the last layer depends on the choice of the loss function, but for the common loss functions of mean-squared error and cross-entropy (with an appropriate choice of activation function) it is simply the difference between the NN output and the label: $\delta^{(L)}=y-t$.

Therefore, to calculate the gradients at each layer we need one vector from the forward pass through the network (the activations) and one vector from the backward pass (the errors).

We see from Eq.~\eqref{eq: backprop} that the error backpropagation consists of two operations. First we must perform an MVM, mirroring the feedforward linear operation \eqref{eq: mvm}. In an ONN, this can be done by light that propagates backwards through the same linear optical arrangement~\cite{wagner1987multilayer}.
The second operation consists in modulation of the MVM output by the activation function derivative and poses a significant challenge for optical implementation. This is because most optical media exhibit similar properties for forward and backward propagation. On the other hand, our application requires an optical element that is (1) nonlinear in the forward direction, (2) linear in the backward direction and (3) modulates the backward light amplitude by the derivative of the forward activation function.

We have solved this challenge with our optical backpropagation protocol, which calculates the right-hand side of Eq.~\eqref{eq: backprop} entirely optically with no opto-electronic conversion or digital processing. The first component of our solution is the observation that many optical media exhibit nonlinear properties for strong optical fields, but are approximately linear for weak fields. Hence, we can satisfy the conditions (1) and (2) by maintaining the back-injected beam at a much lower intensity level than the forward. Furthermore, there exists a set of nonlinear phenomena that also addresses the requirement (3). An example is saturable absorption (SA). The transmissivity of SA medium in the backward direction turns out to approximate the derivative of its intensity-dependent transmission in the forward direction $g'\left(z^{(i-1)}\right)$. This approximation is valid up to a certain numerical factor and only for small values of $ z^{(i-1)}$; however, as shown in our prior work \cite{guo2021backpropagation}, this is sufficient for successful training.

\section*{Multilayer ONN}

Our ONN as shown in Fig.~\ref{fig1}(a,b) is implemented in a free-space tabletop setting. The neuron values are encoded in the transverse spatial structure of the propagating light field amplitude. Spatial light modulators are used to encode the input vectors and weight matrices. The NN consists of two fully-connected linear layers implemented with optical MVM following our previously demonstrated experimental design \cite{spall2020fully}.

This design has a few characteristics that make it suitable for use in a deep neural network. First, it is reconfigurable, so that both neuron values and network weights can be arbitrarily changed. Second, multiple MVM blocks can be cascaded to form a multilayer network, as the output of one MVM naturally forms the input of the next MVM. Using a coherent beam also allows us to encode both positive- and negative-valued weights. Finally, the MVM works in both directions, meaning the inputs and outputs are reversible, which is critical for the implementation of our optical backpropagation algorithm. The hidden layer activation between the two layers is implemented optically by means of SA in a rubidium atomic vapour cell [Fig.~\ref{fig1}(c)].

\section*{Results}

\begin{figure*}[!htb]
    \includegraphics[width=0.9\textwidth]{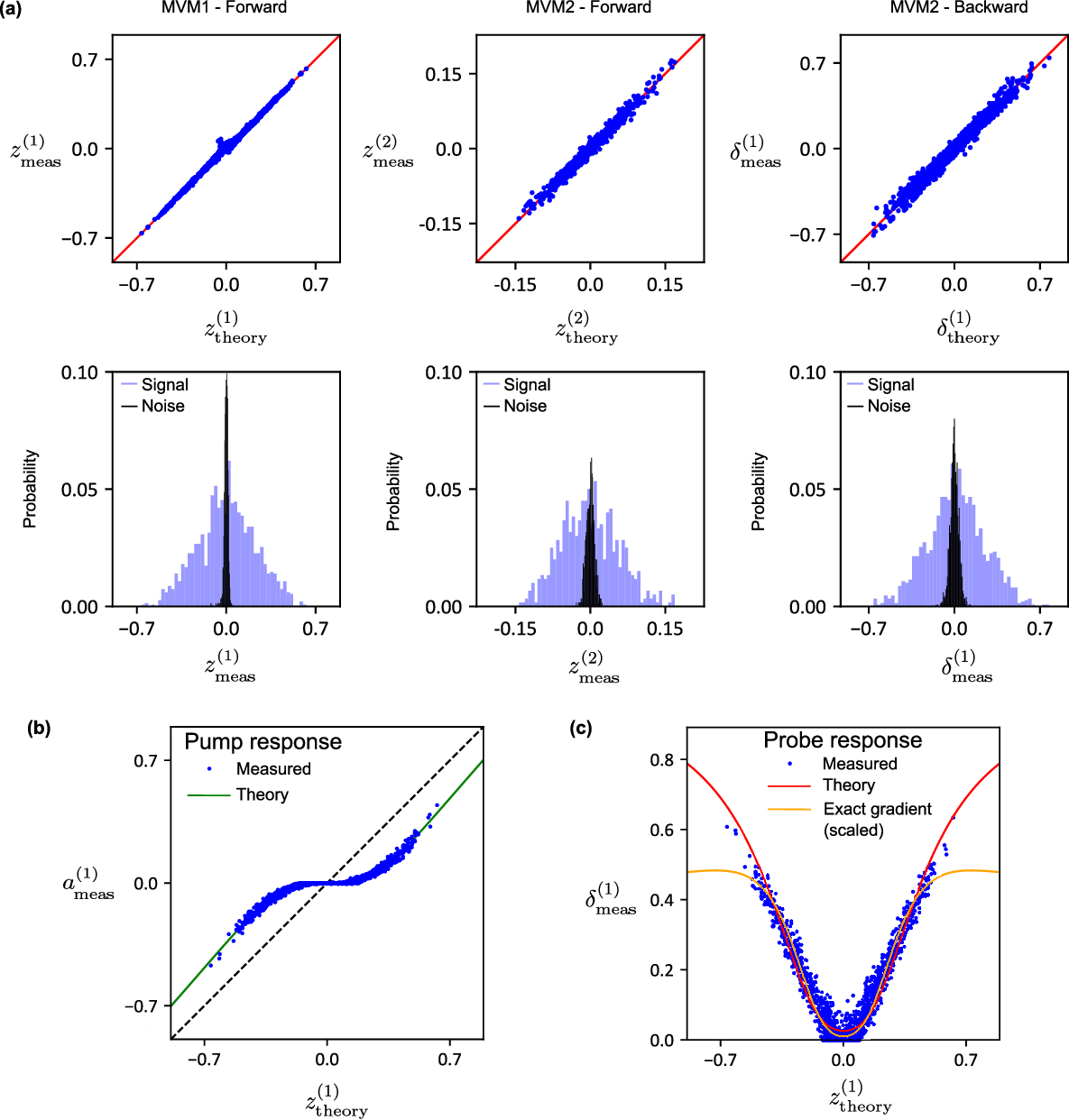}
    \caption{\textbf{Multi-layer ONN characterisation.} \textbf{(a)} Scatter plots of measured against theory results for MVM-1 (first layer forwards), MVM-2a (second layer forwards) and MVM-2b (second layer backwards). All three MVM results are taken simultaneously. Histograms of the signal and noise error for each MVM are displayed underneath. \textbf{(b)} First-layer activations $a^{(1)}_{\rm meas}$ measured after the vapor cell, plotted against the theoretically expected linear MVM-1 output $z^{(1)}_{\rm theory}$ before the cell. The green line is a best fit curve of the theoretical SA nonlinear function. \textbf{(c)} The amplitude of a weak constant probe passed backwards through the vapor cell as a function of the pump $z^{(1)}_{\rm theory}$, with constant input probe. Measurements for both forward and backward beams are taken simultaneously.}
    \label{fig2}
\end{figure*}

\subsection*{Linear layers}
We first set up the linear layers that serve as the backbone of our ONN, and we make sure that they work accurately and simultaneously in both directions --- a highly challenging task that has never been achieved before to our best knowledge.

This involves three MVMs: first layer in the forward direction (MVM-1), second layer in both forward (MVM-2a) and backward (MVM-2b) directions. To characterise these MVMs, we apply random vectors and matrices and simultaneously measure the output of all three: the results for 300 random MVMs are presented in Fig.~\ref{fig2}(a).
To quantify the MVM performance, we define the signal-to-noise ratio (SNR, see Methods for details). As illustrated by the histograms, MVM-1 has the greatest SNR of 14.9, and MVM-2a has a lower SNR of 7.1, as a result of noise accumulation from both layers and the reduced signal range. MVM-2b has a slightly lower SNR of 6.7, because the optical system is optimized for the forward direction. Comparing these experimental results with a simple numerical model, we estimate $1.3\%$ multiplicative noise in our MVMs, which is small enough not to degrade the ONN performance~\cite{spall2022hybrid}.

\subsection*{Nonlinearity}
With the linear layers fully characterized, we now measure the response of the activation units in both directions. With the vapor cell placed in the setup and the laser tuned to resonance with the atomic transition, we pass the output of MVM-1 through the vapor cell in the forward direction. The response as presented in Fig.~\ref{fig2}(b) shows strong nonlinearity. We fit the data with the theoretically expected SA transmissivity (see Supplementary for details), thereby finding the optical depth to be $\alpha_0 = 7.3$, which is sufficient to achieve high accuracy in ONNs~\cite{guo2021backpropagation}. The optical depth and the associated nonlinearity can be easily tuned to fit different network requirements by controlling the temperature of the vapor cell. In the backward direction, we pass weak probe beams through the vapor cell and measure the output. Both the forward and backward beams are simultaneously present in the vapor cell during the measurement.

In Fig.~\ref{fig2}(c) we measure the effect of the forward amplitude $z^{(1)}$ on the transmission of the backward beam through the SA. The theoretical fit for these data --- the expected backward transmissivity calculated from the physical properties of SA --- is shown by the red curve. For comparison, the orange curve shows the rescaled exact derivative $g'\left(z^{(1)}\right)$ of the SA function, which is the dependence required for the calculation \eqref{eq: backprop} of the training signal. Although the two curves are not identical, they both match the experimental data for a broad range of neuron values generated from the random MVM, hence the setting is appropriate for training.

\subsection*{All optical classification}

\begin{figure*}[t]
    \includegraphics[width=\textwidth]{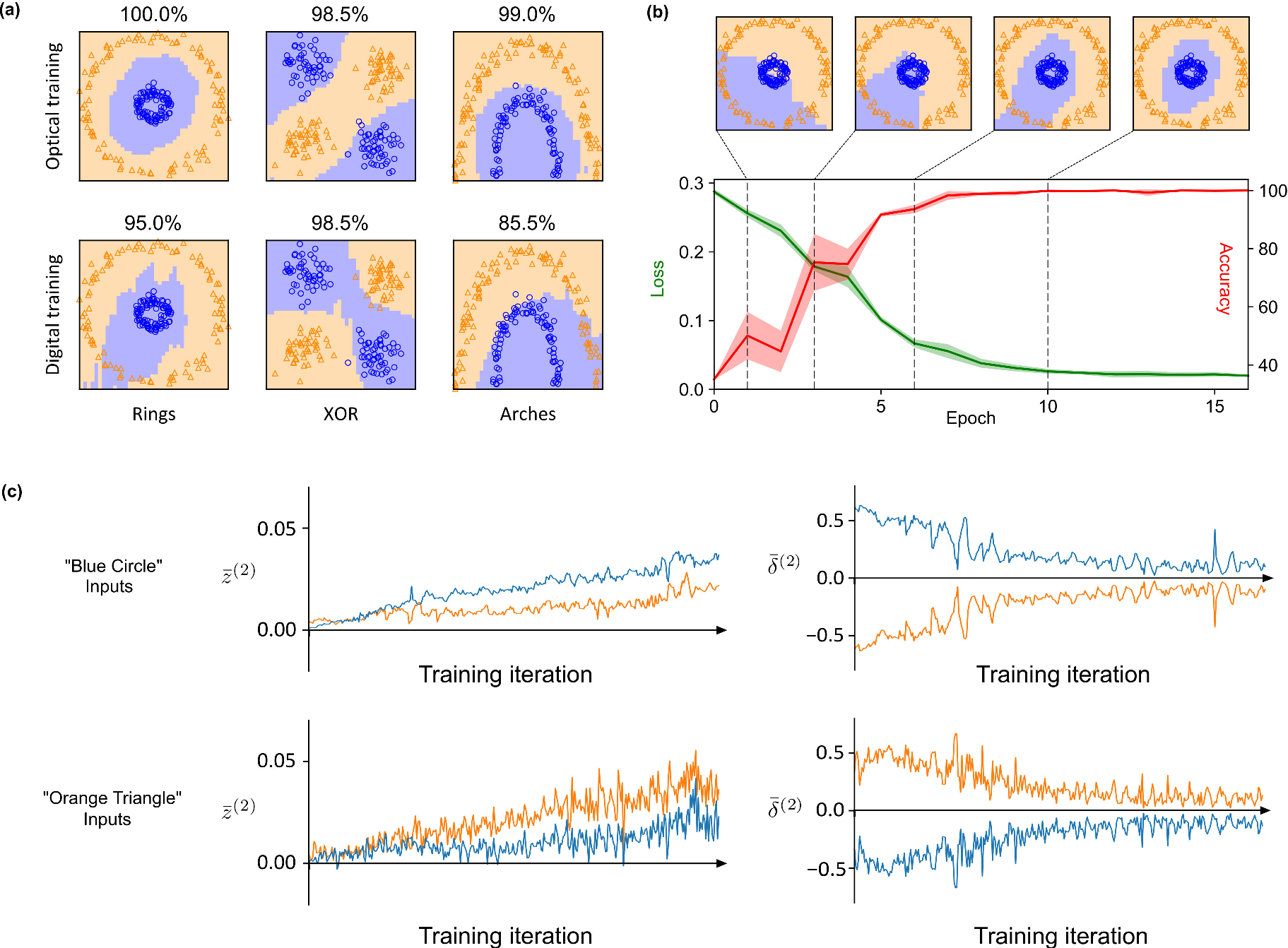}
    \caption{\textbf{Optical training performance.} \textbf{(a)} Decision boundary charts of the ONN inference output for three different classification tasks, after the ONN has been trained optically (top) or \emph{in-silico} (bottom). \textbf{(b)} Learning curves of the ONN for classification of the `rings' dataset, showing mean and standard deviation of the validation loss and accuracy averaged over 5 repeated training runs. Shown above are decision boundary charts of the ONN output for the test set, after different epochs. \textbf{(c)} Evolution of output neuron values, and of output errors, for the training set inputs of the two classes.
    }
    \label{fig3}
\end{figure*}

After setting up the two-layer ONN, we perform end-to-end optical training and inference on classification tasks: distinguishing two classes of data points on a two-dimensional plane (Fig.~\ref{fig3}). We implement a fully-connected feed-forward architecture, with three input neurons, five hidden layer neurons and two output neurons (Fig.~\ref{fig1}). Two input neurons are used to encode the input data point coordinates $(x_1, x_2)$, and the third input neuron of constant value is used to set the first layer bias. The class label is encoded by a `one-hot' vector $(0, 1)$ or $(1, 0)$, and we use categorical cross-entropy as the loss function.

We optically train the ONN on three 400-element datasets with different nonlinear boundary shapes, which we refer to as `rings', `XOR' and `arches' [Fig.~\ref{fig3}(a)]. Another 200 similar elements of each set are used for validation, i.e.~to measure the loss and accuracy after each epoch of training. The test set consists of a uniform grid of equally-spaced $(x_1, x_2)$ values. The optical inference results for the test set are displayed in Fig.~\ref{fig3}(a) by light purple and orange backgrounds, whereas the  blue circles and orange triangles show the training set elements.

For all three datasets, each epoch consists of 20 mini-batches, with a mini-batch size of 20, and we use the Adam optimizer to update the weights and biases from the gradients. We tune hyperparamters such as learning rate and number of epochs to maximise network performance. Table~\ref{table2} summarises the network architecture and hyperparameters used for each dataset.

 \begin{table*}[!htb]

\begin{tabular}{|l|c|c|c|c|c|c|c|}
\hline
\multicolumn{1}{|c|}{Dataset} & \begin{tabular}[c]{@{}c@{}}Input \\ neurons\end{tabular} & \begin{tabular}[c]{@{}c@{}}Hidden\\ neurons\end{tabular} & \begin{tabular}[c]{@{}c@{}}Output\\ neurons\end{tabular} & \begin{tabular}[c]{@{}c@{}}Learning \\ rate\end{tabular} & Epochs & \begin{tabular}[c]{@{}c@{}}Batches \\ per epoch\end{tabular} & Batch size          \\
\hline
Rings & \multirow{3}{*}{2} & \multirow{3}{*}{5} & \multirow{3}{*}{2}  & 0.01  & 16  & \multirow{3}{*}{20}        & \multirow{3}{*}{20} \\ \cline{1-1} \cline{5-6} XOR  & & & & 0.005 & 30  &  & \\ \cline{1-1} \cline{5-6}
Arches&   & &  & 0.01 & 25  &   &\\
\hline
\end{tabular}

\caption{\textbf{Summary of network architecture and hyperparameters used in optical and digital training.}}
\label{table2}
\end{table*}

Figure \ref{fig3}(b) shows the optical training performance on the `rings' dataset. We perform five repeated training runs, and plot the loss and accuracy for the validation set after each epoch of training. To visualise how the network is learning the boundary between the two classes, we also run a test dataset after each epoch. Examples of the network output after 1, 3, 6 and 10 epochs are shown. We see that the ONN quickly learns the nonlinear boundary and gradually improves the accuracy to $100\%$. This indicates a strong optical nonlinearity in the system and a good gradient approximation in optical backpropagation. Details of the training procedure are provided in the Methods section, and results for the other two datasets in Supplementary Note 3.

To better understand the optical training process, we explore the evolution of the output neuron and error vector values in Fig.~\ref{fig3}(c). First, we plot the mini-batch mean value of each output neuron, $\bar{z}_j^{(2)}$, for the inputs from the two classes separately in the upper and lower panels, over the course of the training iterations. We see the output neuron values diverge in opposite ways for the two classes, such that the inputs can be distinguished and correctly classified.

Second, we similarly plot the evolution of the mini-batch mean output error, $\bar{\delta}^{(2)}$, for each neuron. This is calculated as the difference between the network output vector  $a^{(2)}$ and the ground truth label $y$, averaged over each mini-batch. As expected, we see the output errors converge towards zero as the system learns the correct boundary.

\subsection*{Optical training vs \textit{in-silico} training}

To demonstrate the optical training advantage, we  perform \textit{in-silico} training of our ONN as a comparison. We digitally model our system with a neural network of the equivalent architecture, including identical learning rate, number of epochs and all other hyperparamters. The hidden layer nonlinearity and the associated gradient are given by the best fit curve and theoretical probe response of Fig.~\ref{fig2}(b). The trained weights are subsequently used for inference with our ONN. The top and bottom rows in Fig.~\ref{fig3}(a) plot the network output of the test boundary set, after the system has been trained optically and digitally, respectively, for all three datasets. In all cases, the optically trained network achieves almost perfect accuracy, whilst the digitally trained network is clearly not optimised, with the network prediction not matching the data. This is further evidence for the already well-documented advantages of hardware-in-the-loop training schemes.

\section*{Discussion}

\begin{table*}[!htb]
\begin{tabular}{|c|c|c|}
\hline
\textbf{Network layer} & \textbf{Function}  & \textbf{Implementation examples}  \\
\hline
 & MVM & Free-space optical multiplier, photonic crossbar array \\
Linear layer & Diffraction &  Programmable optical mask \\
 & Convolution & Lens Fourier transform \\
 \hline
 & Saturable absorption & Atomic vapour cell, semiconductor absorber, Graphene \\
Nonlinear layer & Saturable gain & EDFA, SOA, Raman amplifier \\
 & Intensity-dependent phase modulation & Optical Kerr medium, thermal nonlinear material\\
 \hline
\end{tabular}
\caption{\textbf{Generalization of the optical training scheme.} EDFA: Erbium-doped fiber amplifier. SOA: semiconductor optical amplifier.}
\label{table}
\end{table*}

Our optical training scheme is surprisingly simple and effective. It adds minimal computational overhead to the network, since it doesn't require \textit{a priori} simulation, or intricate mapping of network parameters to physical device settings. It also imposes minimal hardware complexity on the system, as it requires only a few additional beam splitters and detectors to measure the activation and error values for parameter updates.

Our scheme can be generalised and applied to many other analog neural networks with different physical implementations of the linear and nonlinear layers. We list a few examples in Table.~\ref{table}. Common optical linear operations include MVM, diffraction and convolution. Compatible optical MVM examples include our free-space multiplier and photonic crossbar array~\cite{ohno2022si}, as they are both bidirectional, in the sense that optical field propagating backwards through these arrangements gets multiplied by the transpose of the weight matrix. 
Diffraction naturally works in both directions, hence diffractive neural networks constructed using different programmable amplitude and phase masks also satisfy the requirements \cite{zhou2020situ}. Optical convolution, achieved with the Fourier transform by means of a lens, and mean pooling, achieved through an optical low pass filter, also work in both directions. Therefore, a convolutional neural network can be optically trained as well. Detailed analysis on the generalization to these linear layers can be found in Supplementary Note 4.

Regarding the generalization to other nonlinearity choices, the critical requirement is the ability to acquire gradients during backpropagation. Our pump-probe method is compatible with multiple types of optical nonlinearities: saturable absorption,  saturable gain and intensity-dependent phase modulation~\cite{boyd2020nonlinear}. Using saturable gain as the nonlinearity offers the added advantage of loss compensation in a deep network, and using intensity-dependent phase modulation nonlinearity, such as self-lensing, allows one to build complex-valued optical neural networks with potentially stronger learning capabilities~\cite{zhang2021optical,guo2021backpropagation,cruz2000reinforcement}.

In our ONN training implementation, some computational operations remain digital, specifically the calculation of the last layer error $\delta^{(2)}$ and the outer product \eqref{otimes} between the activation and error vectors. Both these operations can be done optically if needed \cite{guo2021backpropagation}. The error vector can in many cases be obtained by subtracting the ONN output from the label vector by way of destructive interference \cite{spall2022hybrid}. Interference can also be utilized to compute the outer product by fanning out the two vectors and overlapping them in a criss-cross fashion onto a pixelated photosensor.

Our optically trained ONN can be scaled up to improve computing performance. Previously, using a similar experimental setup, we have demonstrated an ONN with 100 neurons per layer and high multiplier accuracy~\cite{spall2022hybrid}, and 1000 neurons can be supported by commercial SLMs or liquid crystal displays. Optical encoders and detectors can work at speeds up to 100 GHz using off-the-shelf components, enabling ultra-fast end-to-end optical computing at low latency. Therefore computational speeds up to $10^{17}$ operations per second are within reach, and our optical training method is compatible with this productivity rate.

\section*{Methods}

\subsection*{Multilayer ONN}

To construct the multi-layer ONN, we connect two optical multipliers in series. For the first layer (MVM-1), the input neuron vector $x$ is encoded into the field amplitude of a coherent beam using a digital micromirror device (DMD), DMD-1. This is a binary amplitude modulator, and every physical pixel is a micromirror that can reflect at two angles representing 0 or 1. By grouping 128 physical pixels as a block, we are able to represent 7-bit positive-valued inputs on DMD-1, with the input value proportional to the number of binary pixels `turned on' in each block.

Since MVM requires performing dot products of the input vector with every row of the matrix, we create multiple copies of the input vector on DMD-1, and image them onto the $W^{(1)}$ matrix mask --- a phase-only liquid-crystal spatial light modulator (LC-SLM), SLM-1 --- for element-wise multiplication. 
The MVM-1 result $z^{(1)}$ is obtained by summing the element-wise products using a cylindrical lens (first optical `fan-in'), and passing the beam through a narrow adjustable slit to select the zero spatial frequency component. The weights in our ONN are real-valued, encoded by LC-SLMs with 8-bit resolution using a phase grating modulation method that enables arbitrary and accurate field control \cite{Arrizon2007Pixelated}.

The beam next passes through a rubidium vapor cell to apply the activation function, such that immediately after the cell the beam encodes the hidden layer activation vector, $a^{(1)}$. The beam continues to propagate and becomes the input for the second linear layer. Another cylindrical lens is used to expand the beam (first optical `fan-out'), before modulation by the second weight matrix mask SLM-2. Finally, summation by a third cylindrical lens (second optical `fan-in') completes the second MVM in the forward direction (MVM-2a), and the final beam profile encodes $z^{(2)}$.

To read out the activation vectors required for the optical training, we insert beam splitters at the output of each MVM to tap-off a small portion of the beam. The real-valued vectors are measured by high-speed cameras, using coherent detection techniques detailed in Supplementary Note 2.

At the output layer of the ONN we use a digital softmax function to convert the output values into probabilities, and calculate the loss function and output error vector, which initiates the optical backpropagation.

\subsection*{Optical backpropagation}

The output error vector, $\delta^{(2)}$ is encoded in the backward beam by using DMD-2 to modulate a beam obtained from the same laser as the forward propagating beam.
The backward beam is introduced to the system through one of the arms of the beam splitter placed at the output of MVM-2a, and carefully aligned so as to overlap with the forward beam. SLM-2 performs element-wise multiplication by the transpose of the second weight matrix.

The cylindrical lens that performs `fan-out' for the forward beam, performs `fan-in' for the backward beam into a slit, completing the second layer backwards MVM (MVM-2b). Passing through the vapor cell modulates the beam by the derivative of the activation function, after which the beam encodes the hidden layer error vector $\delta^{(1)}$. Another beam splitter and camera are used to tap off the backward beam and measure the result.

In practice, two halves of the same DMD act as DMD-1 and DMD-2, and a portion of SLM-1 is used to encode the sign of the error vector. A full experiment diagram is provided in Supplementary Note 1.

Each training iteration consists of optically measuring all of $a^{(1)}$, $z^{(2)}$ and $\delta^{(1)}$. These vectors are used, along with the inputs $x=a^{(0)}$, to calculate the weight gradients according to Eq.~\eqref{otimes} and weight updates, which are then applied to the LC-SLMs. This process is repeated for all the mini-batches until the network converges.
\vspace{1em}

\subsection*{SA activation}

The cell with atomic rubidium vapor is heated to 70 degrees by a simple heating jacket and temperature controller. The laser wavelength is locked to the $D_2$ transition at $780$nm.

The power of the forward propagating beam is adjusted to ensure the beam at the vapor cell is intense enough to saturate the absorption, whilst the maximum power of the backward propagating beam is attenuated to approximately $2\%$ of the maximum forward beam power, to ensure a linear response when passing through the cell in the presence of a uniform pump.

In the experiment, the backward probe response does not match perfectly with the simple two-level atomic model, due to two factors.

First, the probe does not undergo $100\%$ absorption even with the pump turnes off. Second, a strong pump beam causes the atoms to fluoresce in all directions, including along the backward probe path. Therefore, the backward signal has a background offset proportional to the forward signal. To compensate for these issues, three measurements are taken to determine the probe response $\delta^{(1)}$ for each training iteration: pump only; probe only; and both pump and probe. In this way, the background terms due to pump fluorescence and unabsorbed probe could be negated.

\medskip

\noindent\textbf{Acknowledgements}
This work is supported by Innovate UK Smart Grant 10043476. X.G. acknowledges support from the Royal Commission for the Exhibition of 1851 Research Fellowship.

\medskip

\noindent\textbf{Author contributions}
X.G. and A.L. conceived the experiment. J.S. carried out the experiment and performed the data analysis. All the authors jointly prepared the manuscript. This work was done under the supervision of A.L.

\medskip

\noindent\textbf{Competing interests} The authors declare no competing interests in this work.

\bibliography{Bibliography}

\end{document}


\title{Supplementary Materials: Training neural networks with end-to-end optical backpropagation}

\author{James Spall}
\affiliation{Clarendon Laboratory, University of Oxford, Parks Road, Oxford,  OX1 3PU, UK}

\author{Xianxin Guo}
\email{xianxin.guo@physics.ox.ac.uk}
\affiliation{Clarendon Laboratory, University of Oxford, Parks Road, Oxford,  OX1 3PU, UK}
\affiliation{Lumai Ltd, Wood Centre for Innovation, Quarry Road, Headington, Oxford, OX3 8SB, UK}

\author{A. I. Lvovsky}
\email{alex.lvovsky@physics.ox.ac.uk}
\affiliation{Clarendon Laboratory, University of Oxford, Parks Road, Oxford,  OX1 3PU, UK}
\affiliation{Lumai Ltd, Wood Centre for Innovation, Quarry Road, Headington, Oxford, OX3 8SB, UK}

\date{\today}

\maketitle

\tableofcontents

\pagebreak
\section{Theory of NN training by backpropagation}
\label{theory}
Here we reproduce the salient equations of the backpropagation training method for reference; a full explanation can be found in e.g.~\cite{lecun2015deep}. We index the neuron vector at each layer with $i, j$ and $k$ respectively.
We encode an input $x_i$, weight matrix $W_{ji}^{(1)}$ and bias $b_{j}^{(1)}$ in the first layer, and weight matrix $W_{kj}^{(2)}$ in the second layer. MVM-1, hidden layer activation and MVM-2 are then described as follows:
\begin{equation}
    z_{j}^{(1)} = \sum_{i=1}^{N} W_{ji}^{(1)} x_{i}    + b_{j}^{(1)};
\end{equation}
\begin{equation}
    a_{j}^{(1)} = g \left( z_{j}^{(1)}\right);
\end{equation}
\begin{equation}
    z_{k}^{(2)} = \sum_{j=1}^{M} W_{jk}^{(2)} a_{j}^{(1)}.
\end{equation}
Note that the bias is applied only in MVM-1 but not in MVM-2.

We consider two types of loss function:
      categorical cross-entropy (CCE)
\begin{align}
    \mathcal{L} = - \sum_k t_k \text{log}\left(a_{k}^{(2)} \right)
\end{align}
and mean-squared error (MSE)
\begin{equation}
    \mathcal{L}
    = \frac12 \sum_k \left|a_{k}^{(2)}-t_k \right|^2
\end{equation}
where $t_k$ is the one-hot label $(0, 1)$ or $(1, 0)$. In the case of CCE, the second-layer activation $a_{k}^{(2)}$ is found by applying the softmax activation function to $z_{k}^{(2)}$:
\begin{equation}
    a_{k}^{(2)} = \frac{e^{z_{k}^{(2)}}}{\sum_k e^{z_{k}^{(2)}}}
\end{equation}
For MSE, there is no second-layer activation: $a_{k}^{(2)} = z_{k}^{(2)} $.

In both cases, the gradients of the loss function can be calculated to be
\begin{equation}
    \frac{\partial \mathcal{L}}{\partial W_{kj}^{(2)}} =  \delta_k^{(2)} \cdot a_{j}^{(1)},
\end{equation}

\begin{equation}
    \frac{\partial \mathcal{L}}{\partial W_{ji}^{(1)}} = \delta_j^{(1)} \cdot x_{i}
\end{equation}
and

\begin{equation}
    \frac{\partial \mathcal{L}}{\partial b_{j}^{(1)}} = \delta_j^{(1)},
\end{equation}

where
\begin{equation}
    \delta_k^{(2)} = \left( a_{k}^{(2)} - t_k \right)
\end{equation}

and
\begin{equation}
    \delta_j^{(1)} = \left( \sum_{k=1}^{L} w_{kj}^{(2)} \delta_{k}^{(2)} \right)  g'\left( z_{j}^{(1)}\right).
\label{eqn:backprop}
\end{equation}
We can therefore directly calculate the second weight matrix update, but to calculate the updates to the first layer, we use our optical backpropagation scheme to optically perform the calculation \eqref{eqn:backprop}.
\newpage
\section{Experimental setup}
\label{experiment-setup}
\begin{figure*}[b]
    \includegraphics[height=0.8\textheight]{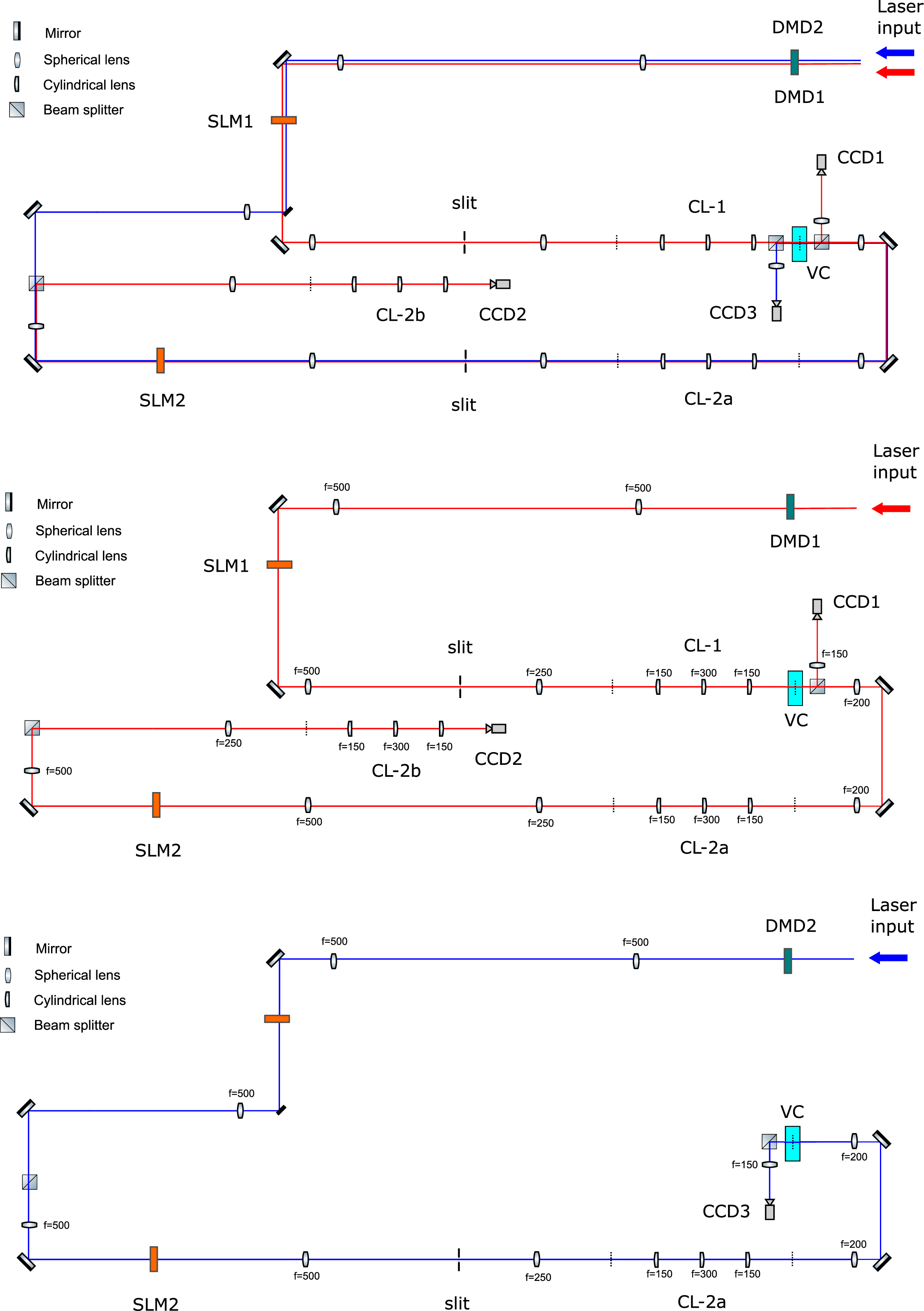}
    \caption{\textbf{(a)} Full experiment schematic. The forward beam is shown in red, and backward beam in blue. The respective paths are shown individually in \textbf{(b)} and \textbf{(c)}. Two sections of the same DMD are used for the input of both the forward and backward paths. CL-1, CL-2a and CL-2b are the three cylindrical lenses discussed in the main text. VC: vapor cell, CCD: digital camera.}
    \label{fig: full diagram}
\end{figure*}

Figure \ref{fig: full diagram} presents the full diagram of the experimental setup, whose simplified version can be found in Fig.~1 of the main text.
\newpage

Each fan-in and fan-out process is performed by a set of three cylindrical lenses, equally spaced at $f=150$ mm between the input and output planes. The central lens, of focal length $2f=300$ mm, performs the Fourier transform of the field in the input plane into the output plane with respect to one of the transverse coordinates. The other two cylindrical lenses, oriented at $90^\circ$ with respect to the central one, have the focal length $f=150$ mm and implement $4f$ imaging of the input plane to the output plane with respect to the other transverse coordinate, preventing unwanted diffraction and maintaining the correct phase and amplitude profile.
The sets CL-1 and CL-2a perform a Fourier transform of the horizontal dimension while imaging the vertical dimension. The set CL-2b does the opposite. Note that because the ONN input in both directions is supplied in the `fanned-out' state, no Fourier transformation is required between the forward input and SLM-1 as well as between the backward input and SLM-2; the inputs are simply imaged onto the SLMs.

Our ONN uses a coherent beam to encode real-valued elements. Below we describe how we encode and detect negative numbers.

The forward ONN input contains the test set element coordinates, which are always positive and can hence be encoded in a single DMD. On the other hand, the errors that constitute the backward ONN input can be negative. We encode their absolute value by means of the DMD, and subsequently  apply a $\pi$ phase shift to encode negative numbers by reflecting the beam from a dedicated section of SLM-1.

The weight matrices are encoded by means of a phase grating on the phase-only LC-SLMs. By spatially varying the offset and height of the grating, we can control both the amplitude and phase of each weight matrix element. The pairwise matrix-vector products are then emitted into the first diffraction order, which is isolated by a spatial filter after reflection from the LC-SLM \cite{spall2020fully,spall2022hybrid}.

We measure the sign of the fields in the two layers by means of homodyne detection, similar to our previous work \cite{spall2022hybrid}. Dedicated region on DMD-1 create two uniform reference beams, which pass through the system along with the MVM signal. The reference beam for the detection of the MVM-1 output is displaced horizontally with respect to the signal beam, whereas that for the MVM-2 output is displaced vertically. In this way, the second reference beam does not interfere with the signal until the very last cylindrical lens set, which acts to perform the MVM summation as well as converge and mix the signal and second reference beam. For the second layer, the reference is always brighter than the signal, and because the two beams follow the same path they are phase-stable. Therefore intensity measurement of the interfered beams gives $I_{\text{meas}} = |E_{\text{sig}} + E_{\text{ref}}|^2$ and the MVM result can be simply read out as $E_{\text{sig}} = \sqrt{I_{\text{meas}}} - E_{\text{ref}}$. The second reference beam passes through the activation vapor cell spatially separated from the signal, so experiences some attenuation but remains uniform, and does not interfere with the pump-probe process.

In contrast, the first reference beam does overlap with the signal in the first slit, due to the action of CL-1, and they subsequently propagate together through the vapor cell. After the cell, the overlapping beams are tapped off via a beam splitter for measurement. Two measurements are performed sequentially. First, the signal intensity $I_{\text{meas1}} = \left|E_{\text{sig}}\right|^2$ without the reference beam is measured, yielding the absolute value $\left|E_{\text{sig}}\right| = \sqrt{I_{\text{meas1}}}$. Second, the reference beam is turned on and the intensity $I_{\text{meas2}} = |E_{\text{sig}} + E_{\text{ref}}|^2$ is detected, telling us whether $E_{\text{sig}}$ is positive or negative (in contrast to MVM-2, the first reference beam is significantly weaker than the typical signal). 
The two measurements are necessary because the combination of the signal and reference beams is affected by the SA nonlinearity in a nontrivial fashion.

\pagebreak
\section{Training XOR and Arches}
\label{xor_arches}
\begin{figure*}[!htb]
    \includegraphics[width=0.5\textwidth]{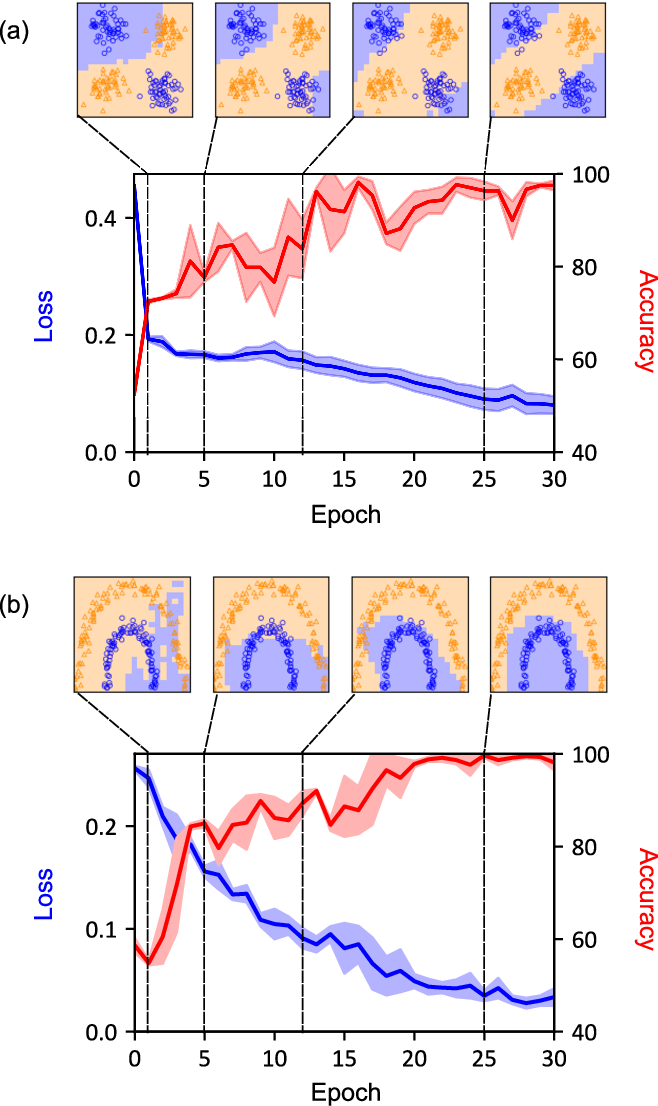}
    \caption{\textbf{(a)} Mean and standard deviation of validation loss and accuracy during optical training of the `XOR' dataset. Shown above are example boundary plots of the test dataset after certain epochs. \textbf{(b)} As above, for the `arches' dataset.}
    \label{fig: xor_rainbow_result}
\end{figure*}

Figure \ref{fig: xor_rainbow_result} shows the learning curves and the decision boundary evolution for the two training sets, akin to Fig.~3(a) in the main text.

\pagebreak

\section{Optical backpropagation through the linear layer}

To carry out optical backpropagation in an ONN, the linear layers need to be bidirectional such that light can propagate in both directions, and achieve a transpose of the weighting or transform matrix in the backward direction. In this section, we will explain how to achieve optical backpropagation through three types of linear layers: MVM, diffraction and convolution layer.

\subsection{Photonic MVM}
Our optical MVM is built with free-space optics using lenses and spatial light modulators following the `Stanford multiplier' design first proposed by Goodman in 1978~\cite{goodman1978fully}. There are of course different ways to achieve optical MVM. For example, on the photonic platform, one can also perform MVM using a crossbar array architecture, as illustrated in Fig.~\ref{fig: LinearLayers}(a). In this setup, a micro-resonator is placed at each crossing of two waveguides. Forward propagating signal enters from the left side of the crossbar array and exits from the bottom. Each element of the weight matrix is controlled by a variable beamsplitter --- for example, a micro-resonator, so the transmission and reflection of the signal at each crossing depends on the resonance condition, which can be electrically controlled. Forward signals from multiple rows are thus weighted and summed up to yield the MVM output. To perform optical backpropation, we can inject the error vector from the bottom. The microresonators will then reflect it towards the left side, such that the signal and error beams counter-propagate. A more convenient option is to send the error vector from the top side of the crossbar array and direct it towards the right output ports~\cite{ohno2022si}. The error vector from different columns are weighted and summed up, yielding MVM result with a transpose of the weight matrix. The signal and error beams should be separated either spectrally or temporally, such that the cross-talks or undesired interference at the output ports can be minimized.

Similar backpropagation is possible in other photonic implementations of MVM, such as that based on networks of Mach-Zehnder interferometers (MZIs). However, the crossbar array case is simpler in that the reflectivity of every beamsplitter directly corresponds to an element of the weight matrix. Hence the backpropagation process as described in this paper will directly yield the gradient for the beamsplitters update. In the case of MZI network, the relation between the weight matrix and the beamsplitter reflectivities is complicated. An additional calculation is therefore needed to obtain the gradient of each phase shifter setting. As shown recently shown both theoretically~\cite{hughes2018training} and experimentally~\cite{pai2023experimentally}, this additional computing step can also be implemented optically.

\subsection{Diffraction layer}
\begin{figure*}[t]
    \includegraphics[width=\textwidth]{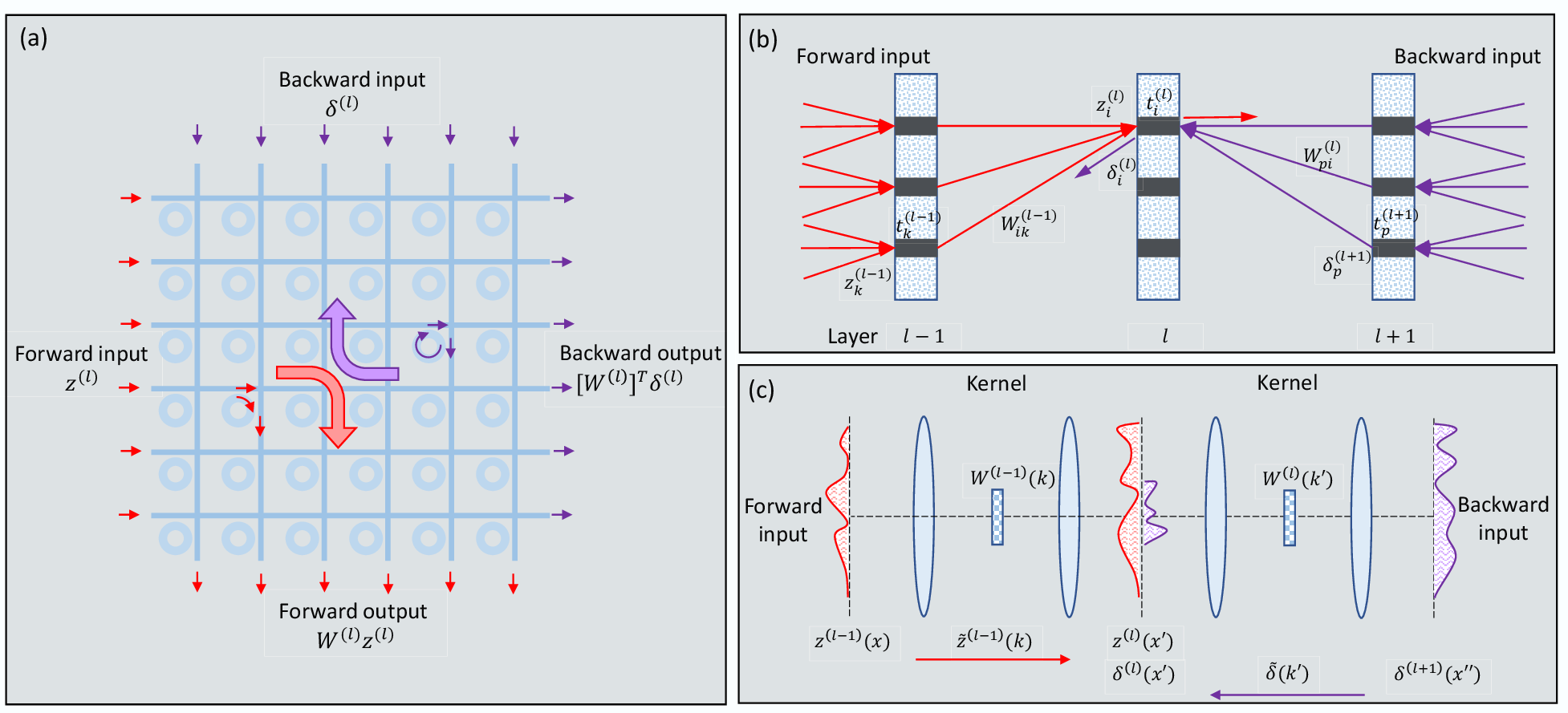}
    \caption{\textbf{Different linear optical layers supporting direct optical backpropagation.} \textbf{(a)} A photonic crossbar MVM built with waveguides and micro-resonators. \textbf{(b)} A diffractive neural network constructed with multiple phase or amplitude masks. \textbf{(c)} Optical convolution layers implemented via 4f lens configurations. }
    \label{fig: LinearLayers}
\end{figure*}

In recent years, diffractive neural networks have attracted significant research interest~\cite{lin2018all}, and they have been shown to achieve high performance in various machine learning and machine vision tasks such as image recognition and object tracking. Diffraction is a natural optical phenomenon, and diffractive layers can be constructed using 3D printed masks, liquid-crystal SLMs and meta-materials. In Fig.~\ref{fig: LinearLayers}(b) we depict the operation principle of a diffractive neural network. In this network, neurons at neighbouring layers are connected via diffraction, and we denote $W_{ki}^{(l-1)}$ as the complex factor acquired by the amplitude of the light field propagating from neuron $z_{k}^{(l-1)}$ at layer $l-1$ to neuron $z_{i}^{(l)}$ at layer $l$. These coefficients are fixed by the ONN geometry and usually not trainable once the network is fabricated. Trainable parameters in the network are the complex transmission coefficients $t_{\cdot}^{(\cdot)}$ of each neuron at different layers.

During the forward direction, signals propagate from one layer to the next as
\begin{equation}
    z_{i}^{(l)}=\sum_{k}W_{ik}^{(l-1)}t_{k}^{(l-1)}z_{k}^{(l-1)}.
    \label{eq: diffractive FW}
\end{equation}
To train the network, we calculate the gradient of the loss function with respect to the transmission coefficient of each layer:
\begin{align}\nonumber
    \frac{\partial \mathcal{L}}{\partial t_{i}^{(l)}} &= \sum_p{\frac{\partial \mathcal{L}}{\partial z_{p}^{(l+1)}}\cdot \frac{\partial z_{p}^{(l+1)}}{\partial t_{i}^{(l)}}}  \\
    &= \sum_p{\frac{\partial \mathcal{L}}{\partial z_{p}^{(l+1)}} W_{pi}^{(l)}z_{i}^{(l)}},
    \end{align}
where $W_{pi}^{(l)}$ is known \emph{a priori}, and $z_{i}^{(l)}$ can be measured experimentally in the forward propagation step. We denote  $\delta_{p}^{(l+1)}=\partial \mathcal{L}/ \partial z_{p}^{(l+1)}$ as the error vector at layer $l+1$, then the error vector at the preceding layer $l$ can also be calculated by applying the chain rule of calculus:
\begin{align} \nonumber
    \delta_{i}^{(l)} &= \sum_{p}{\frac{\partial \mathcal{L}}{\partial z_{p}^{(l+1)}}\cdot  \frac{\partial z_{p}^{(l+1)}}{\partial z_{i}^{(l)}}} \\
    &= \sum_{p}{\delta_{p}^{(l+1)}W_{pi}^{(l)}t_{i}^{(l)}}.
    \label{eq: diffractive BW}
\end{align}
The last line indicates that the error vector at a given layer is calculated by multiplying the error vector at a subsequent layer with the diffractive connection matrix transpose, then element-wise multiplied with the mask transmission coefficients. This is exactly the optical backpropagation process. The physical forward and backward propagation process is better evidenced by re-writing Eq.~\ref{eq: diffractive FW} and Eq.~\ref{eq: diffractive BW} in the matrix form:
\begin{align}
    {z}^{(l)} &= {z}^{(l-1)}\cdot {t}^{(l-1)}\times W^{(l-1)}, \\
    {\delta}^{(l)} &= {\delta}^{(l+1)}\times [W^{(l)}]^T\cdot {t}^{(l)}.
\end{align}

\subsection{Convolution layer}
The convolutional neural network represents a basic type of neural networks, and it is ubiquitous in processing complex machine vision tasks. It can capture key features of images efficiently through the use of convolution kernels layer-by-layer. It is well known that convolution is equivalent to multiplication in the Fourier domain, and Fourier transform can be optically performed with a simple lens.

Figure~\ref{fig: LinearLayers}(c) depicts two cascaded optical convolution layers using two lenses per layer comprising a 4f system. The first lens at layer $l-1$ performs Fourier transform of the incoming signal $z^{(l-1)}(x)$:
\begin{equation}
     \tilde{z}^{(l-1)}(k) = \mathcal{F}[z^{(l-1)}(x)] = \int{z^{(l-1)}(x) e^{-ikx}dx}.
\end{equation}
Then a mask $W(k)$ (which is the Fourier transform of the convolution kernel) placed at the focal plane modulates the signal spectrum $\tilde{z}^{(l-1)}(k)$, and a second lens performs another Fourier transform to complete the convolution:
\begin{align}
    z^{(l)}(x') &= \mathcal{F}[\tilde{z}^{(l-1)}(k)W^{(l-1)}(k)] = \iint{z^{(l-1)}(x)W^{(l-1)}(k) e^{-ik(x+x')}dxdk}.
\end{align}
To update the kernels, we calculate the gradient of the loss function with respect to the transmissivity $W(k)$:
\begin{align}\nonumber
    \frac{\partial\mathcal{L}}{\partial W^{(l-1)}(k)} &= \int{\frac{\partial\mathcal{L}}{\partial z^{(l)}(x')} \frac{\partial z^{(l)}(x')}{\partial W^{(l-1)}(k)} dx'} \\\nonumber
    &= \iint{\frac{\partial \mathcal{L}}{\partial z^{(l)}(x')} z^{(l-1)}(x)e^{-ik(x+x')} dxdx'} \\\nonumber
    &= \iint{\delta^{(l)}(x') z^{(l-1)}(x)e^{-ik(x+x')} dxdx'} \\
    &=\tilde{z}^{(l-1)}(k)\cdot \tilde{\delta}^{(l)}(k).
\end{align}
In the calculation above we have introduced the error vector $\delta^{(l)}(x')=\dfrac{\partial \mathcal{L}}{\partial z^{(l)}(x')}$ and its Fourier spectrum $\tilde{\delta}^{(l)}(k)=\int\delta(x)e^{-ikx'}dx$. 
This result means that the kernel gradient is the dot product of the signal spectrum and error vector spectrum. Next we show how the error vector spectrum can be obtained by applying the chain rule of calculus:
\begin{align}\nonumber
    \tilde{\delta}^{(l)}(k) &= \int{\frac{\partial \mathcal{L}}{\partial z^{(l)}(x')} e^{-ikx'} dx'} \\\nonumber
    &= \iint{\frac{\partial \mathcal{L}}{\partial z^{(l+1)}(x'')} \frac{\partial z^{(l+1)}(x'')}{\partial z^{(l)}(x')} e^{-ikx'} dx'dx''} \\\nonumber
    &= \iiint{\delta^{(l+1)}(x'') W^{(l)}(k')e^{-ik'(x'+x'')}e^{-ikx'}dk'dx'dx''} \\
\nonumber
    &= \int\delta^{(l+1)}(x'') W^{(l)}(k')e^{-ik'x''}dx''\int e^{-i(k'+k)x'} dx' \\
    &= \tilde{\delta}^{(l+1)}(-k) \cdot W^{(l)}(k).
\end{align}
Here we used the fact that $\int e^{-i(k'+k)x'}dx'$ is the Dirac delta function with respect to $k'+k$.

From this result, we see that the error vector spectrum at layer $l$ is obtained by multiplying the error vector spectrum at the next layer $l+1$ with the kernel at layer $l$. The negative sign in front of $k$ in $ \tilde{\delta}^{(l+1)}(-k)$ is due to the image inversion in a 4f system.  Therefore, optical convolution layers also support direct optical backpropagation.

\bibliography{Bibliography}